\documentclass[aps,prl,groupedaddress,twocolumn,10pt]{revtex4-1}
\usepackage{graphicx}
\usepackage{epstopdf}
\usepackage[english]{babel}
\usepackage[T1]{fontenc}
\usepackage{verbatim}
\usepackage{float}

\usepackage{makecell}
\usepackage{color}
\usepackage{multirow}
\usepackage{booktabs}
\usepackage{amsmath}
\usepackage{amssymb}
\usepackage{amsthm}
\usepackage{bbm}
\usepackage{hyperref}
\usepackage[T1]{fontenc}
\usepackage[usenames,dvipsnames]{xcolor}
\hypersetup{colorlinks=true, linkcolor=blue, urlcolor=blue, citecolor=blue}

\usepackage[normalem]{ulem}

\renewcommand{\imath}[0]{\mathsf{i}}

\usepackage[nodayofweek,level]{datetime}

\begin{document}

\title{Cellular uptake of active nonspherical nanoparticles}
\author{Ke Xiao}
\email{xiaoke@ucas.ac.cn}
\affiliation{Department of Physics, College of Physical Science and Technology, Xiamen University, Xiamen 361005, People's Republic of China}
\affiliation{Wenzhou Institute, University of Chinese Academy of Sciences, Wenzhou 325016, People's Republic of China}
\author{Jing Li}
\affiliation{Department of Physics, College of Physical Science and Technology, Xiamen University, Xiamen 361005, People's Republic of China}
\author{Rui Ma}
\affiliation{Fujian Provincial Key Lab for Soft Functional Materials Research, Research Institute for Biomimetics and Soft Matter, Department of Physics, College of Physical Science and Technology, Xiamen University, Xiamen 361005, People's Republic of China}
\author{Chen-Xu Wu}
\email{cxwu@xmu.edu.cn}
\affiliation{Fujian Provincial Key Lab for Soft Functional Materials Research, Research Institute for Biomimetics and Soft Matter, Department of Physics, College of Physical Science and Technology, Xiamen University, Xiamen 361005, People's Republic of China}

\begin{abstract}
Due to the potential applications in biomedical engineering, it becomes more and more important to understand the process of engulfment and internalization of nanoparticles (NPs) by cell membranes.
Despite the fact that the interaction between cell membranes and passive particles has been widely studied, the interaction between cell membranes and self-propelled nonspherical NPs remains to be elucidated.
Here we present a theoretical model to systematically investigate the influence of the active force, aspect ratio of NPs, particle size and membrane properties (adhesion energy density and membrane tension) on the cellular uptake of a nonspherical nanoparticle.
It is found that the active force generated by an NP can trigger a type of first-order wrapping transition from a small partial wrapping state to a large one.
In addition, the phase diagram in the force-aspect ratio (particle size, adhesion energy density and membrane tension) space displays more complex behaviors compared with that for the passive wrapping mediated merely by adhesion.
These results may provide a useful guidance to the study of activity-driven cellular entry of active particles into cells.
\end{abstract}
\date{\today}

\maketitle

\section{INTRODUCTION}
The transport of nano-sized particles across cells or vesicles made of lipid-bilayer membranes is a ubiquitous phenomenon in biological processes with many applications in biomedical and biotechnology fields ranging from drug and gene delivery~\cite{Panyam2003,Xu2018,Wang2019,N.Pardi2015} to biomedical imaging and sensing~\cite{R.Weissleder2006,D.Peer2007,W.Kukulski2012}.
Cellular uptake is a key pathway for transporting cargo into cell via being engulfed and internalized by cell membranes, a process related to the interaction between cell membranes and NPs.
 Such a wrapping process plays an integral role in a wide range of health related aspects~\cite{B.Alberts2005} such as nutrient import, signal transduction, neurotransmission~\cite{F.Frey2019PRE,G.V.Meer2004,N.Walani2015PNAS}, and cellular entry and exit of viruses, pathogens and parasites into host cells~\cite{J.Mercer2010,S.Dasgupta2014BJ,G.Bao2005,S.Zhang2015}. In addition, it is also important for designing diagnostic and therapeutic agents due to the rapid development of NPs for the delivery of, for example, anticancer agents~\cite{M.E.Davis2008}. For example, specifically enveloped particles are utilized to serve as targeting drug delivered into tumor cells~\cite{D.Peer2007,W.Rao2015,Y.Min2015,R.Imani2017,M.Wang2023}, based on the understanding of the interactions between cell membranes and NPs.
Despite its biological importance, it is still not fully understood how the active force and the aspect ratio of NPs and membrane properties (adhesion energy density and membrane tension) affect the wrapping behaviors.

Investigations concerning the engulfment and internalization of passive particles by cell membranes have been extensively conducted experimentally, theoretically, and numerically.
 Among them, many studies are focused on the influence of physical parameters, including particle size~\cite{Deserno2002,Deserno2003,M.Deserno2004,S.Zhang2009,B.D.Chithrani2006,J.Agudo2015,C.Contini2020}, shape~\cite{J.Midya2023,F.Frey2019,K.Yang2010,Z.Shen2019,S.Dasgupta2014,S.Dasgupta2013,A.H.Bahrami2013,D.M.Richards2016,L.P.Chen2016}, elastic properties of invading particles~\cite{J.Midya2023,X.Yi2011,J.C.Shillcock2005,X.Yi2014,A.Verma2010,X.Ma2021}, ligand and receptor density~\cite{H.Yuan2010PRL,H.Yuan2010,T.Wiegand2020}, as well as the mechanical properties of the membrane~\cite{J.Agudo-Canalejo2015, H.T.Spanke2020}, based on adhesion-mediated wrapping mechanism.
In recent years, there has been a growing research interest in the interactions between biological self-propelled bacterial pathogens (\textit{Rickettsia rickettsii} or \textit{Listeria monocytogenes}, \textit{Escherichia coli} bacteria, and \textit{Bacillus subtilis} bacteria, etc.) or synthetic self-propelled particles (synthetic Janus particles) and the cell membranes~\cite{H.R.Vutukuri2020,S.C.Takatori2020,C.Wang2019,Y.Li2019,M.S.E.Peterson2021,L.LeNagarda2022}.
One of the main features of these self-propelled bacterial pathogens or synthetic artificial particles is that they are able to generate mechanical forces by consuming energy from their environment, which often results in motion~\cite{S.Ramaswamy2010}.
For instance, it has been found that \textit{Rickettsia rickettsii} are able to produce active force to facilitate their mobility by forming actin tails~\cite{P.M.Colonne2016}, and \textit{Listeria monocytogenes} can generate active force to push out a tube-like protuberance from the plasma membrane by hijacking the actin polymerization-depolymerization apparatus of their host~\cite{J.A.Theriot1992,J.R.Robbins1999,T.Chakraborty1999,F.E.Ortega2019,G.C.Dowd2020}. Furthermore, the interplay of self-propelled particles with cell membranes also leads to rich intriguing dynamical behaviors and functions such as membrane fluctuations and large deformations~\cite{H.R.Vutukuri2020,S.C.Takatori2020}, shape transformations~\cite{C.Wang2019,Y.Li2019,M.S.E.Peterson2021}, and even deformation of lipid vesicles into flagellated swimmers~\cite{L.LeNagarda2022}.
The specific interactions between membranes and self-propelled bacterial pathogens or artificial self-propelled particles plays a key role in designing active matter systems~\cite{A.T.Brownet2016}. How the active force of these self-propelled agents affects the wrapping behaviors remains to be elucidated.

Nevertheless, the studies mentioned above are confined within spherical NPs. In many biological systems, active NPs such as the wrapped pathogens or viruses, can be highly nonspherical~\cite{S.Dasgupta2014,C.Hulo2011}, such as egg-shaped malaria parasite~\cite{S.Dasgupta2014} and cylindrical Listeria monocytogenes. Moreover, the shape of the wrapped particles also affects the wrapping behaviors in cellular uptake, and scientists have been motivated to model the wrapping behaviors of artificial particles with various geometries such as ellipsoids, rod-like particles, and capped cylinders~\cite{J.Midya2023,F.Frey2019,K.Yang2010,Z.Shen2019,S.Dasgupta2014,S.Dasgupta2013,A.H.Bahrami2013,D.M.Richards2016,J.A.Champion2006,N.Doshi2010,D.Paul2013}.
However, in these studies, particle activity was not taken into account.

Theoretically, the interplay of a lipid membrane with an NP is typically governed by only a few physical parameters (membrane bending rigidity, membrane tension, and adhesion energy density), through which the membrane resists bending and stretching. The deformation of a membrane can also occur as a consequence of adhesive interactions between the membrane and the particle, characterized by an adhesion energy per unit area. A detailed and comprehensive investigation of how the wrapping behaviors depend on the active force, the particle's aspect ratio, and the membrane properties (adhesion energy density and membrane tension) is needed.

To model the action of forces on a membrane, we adopted the spirit of continuum mechanics by treating the membrane as a smooth surface and incorporating the work done by the force into the total energy of the membrane~\cite{Brochard2006,Derenyi2002,B.Bozic1997}. To determine the equilibrium shape of the membrane, the corresponding variational problem carried out here is mathematically equivalent to many of previous papers~\cite{Derenyi2002,B.Bozic1997,B.Sabass2016}.

In this work, we use energy minimization to calculate and predict shapes and wrapping states for an ellipsoidal NP at an initially flat membrane.
Our article is organized as follows. In Section II, we describe our theoretical model including the numerical method employed and the parameters we used. Section III is devoted to results and discussions, including the influence of the active force on the wrapping states of ellipsoid and its corresponding phase diagram, the effects of the particle's aspect ratio and the membrane properties and its corresponding phase diagrams. Section IV is devoted to conclusion.

\section{THEORETICAL MODEL}
\begin{figure}[htp]
  \includegraphics[width=\linewidth,keepaspectratio]{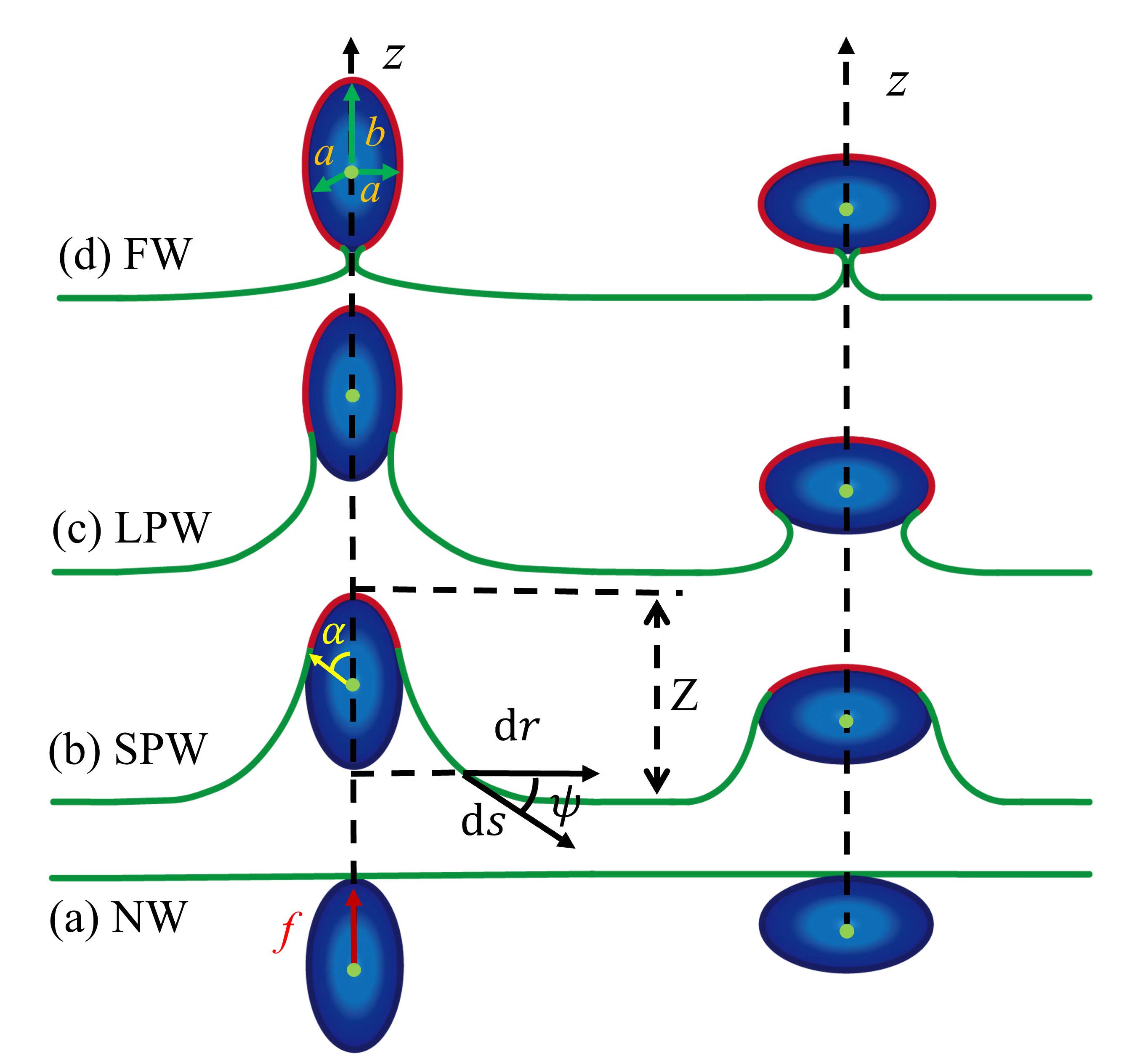}
  \caption{(Color online) Schematic of the four different wrapping states: (a) nonwrapping (NW), (b) small partial wrapping (SPW), (c) large partial wrapping (LPW), and (d) full wrapping (FW). The axisymmetric parameterization of the membrane shape is shown in (b). Here, $\alpha$ denotes the polar angle of the point where the membrane detaches from the NP surface, and we take the wrapping angle $\alpha$ as an order parameter, and define an SPW state if the wrapping degree is shallow, and an LPW state if the wrapping degree is deep~\cite{S.Dasgupta2014}.\label{uptakeprocess}}
\end{figure}

We consider an initially flat membrane pushed by an active self-propelled rigid ellipsoidal NP (prolate or oblate spheroid) with its principle rotational axis orthogonal to the membrane, as shown in Fig.~\ref{uptakeprocess}. For simplicity, we assume that the active force is constant and always falls strictly along the $z$ direction in a way so that the system obeys rotational symmetry and the particle will not rotate during the wrapping process. Here it should be noted that for a passive particle, it may undergo orientational rotation, possibly due to stochastic thermal fluctuation of the membrane~\cite{L.P.Chen2016}.
The shape of this ellipsoidal particle can be defined by the shape equation in Cartesian coordinates $x$, $y$, $z$,
\begin{align}
\frac{x^2+y^2}{a^2}+\frac{z^2}{b^2}=1,\label{Ellipsoidshape}
\end{align}
where $a$ and $b$ denote the semi-axes perpendicular to and along the principle rotational axis, respectively.
The geometry of the particle is parameterized by the aspect ratio $e=b/a$, with $e<1$ for oblate ellipsoids, $e=1$ for spheres, and $e>1$ for prolate ellipsoids, respectively.
Using the classical Canham-Helfrich continuum membrane model~\cite{Helfrich1973,F.Julicher1994,M.Deserno2004}, the total free energy of such a system is given by
\begin{align}
E^{\mathrm{tot}}=\int_{A_{\rm mem}}\frac{\kappa}{2}(2H)^2 dA +\sigma\Delta A - \int_{A_{\rm ad}}\omega~dA - fZ,\label{HelfrichFE}
\end{align}
where the first term is contributed by the bending energy of the membrane, with $\kappa$ the bending rigidity and $H$ the local mean curvature.
The second term is the tension energy, with $\sigma$ being the membrane tension and $\Delta A$ being the excess area induced by wrapping. The third term represents the gain in adhesive energy, characterized by a negative adhesion energy per unit area $-\omega$. The deformation of the membrane can not only be induced by the adhesive interactions between the membrane and the particle, but also occurs as a consequence of the work done by the active force, as represented by the last therm of Eq.~(\ref{HelfrichFE}). Here the membrane is assumed to be pushed by the active force $f$ to a height of $Z$ [see Fig.~\ref{uptakeprocess}].

An ellipsoidal NP-membrane system can be divided into two parts: the wrapped part and the free part of the membrane.
In practice, it is convenient use the parametric equations $x=a\sin\theta\cos\phi$, $y=a\sin\theta\sin\phi$ and $z=b\cos\theta$, to describe the ellipsoidal NP surface. Given this, the area element can be obtained in terms of polar angle: $dA=2\pi a \sin\theta\sqrt{a^2\cos^2\theta+b^2\sin^2\theta}~d\theta$, and the energy generated by the particle-membrane adhesive interaction is written as
\begin{align}
E_{\rm ad}=-\int_0^{\alpha}2\pi\omega a \sin\theta \sqrt{a^2\cos^2\theta+b^2\sin^2\theta}~d\theta, \label{E_adhesion}
\end{align}
where $\alpha$ is the wrapping angle. The mean curvature of the ellipsoidal particle can be calculated as
\begin{align}
H=\frac{c_1+c_2}{2}=\frac{b[2a^2+(b^2-a^2)\sin^2\theta]}{2a[a^2+(b^2-a^2)\sin^2\theta]^{3/2}},\label{Meancurvature}
\end{align}
where $c_1$ and $c_2$ are the two principle curvatures. Hence, the bending energy of the adhesive (wrapping) part can be written as
\begin{align}
E_{\rm bend}^{\rm ad}=&\int_0^{\alpha}\pi\kappa \sin\theta\frac{b^2[2a^2+(b^2-a^2)\sin^2\theta]^2}{a[a^2+(b^2-a^2)\sin^2\theta]^3}\times \notag\\ &\sqrt{a^2\cos^2\theta+b^2\sin^2\theta}~d\theta.\label{E_bend}
\end{align}
Similarly, the contribution made by the surface tension of the adhesive (wrapping) part can be given by
\begin{align}
E_{\rm ten}^{\rm ad}=&\int_0^{\alpha}2\pi\sigma a \sin\theta \biggl(1-\frac{a\cos\theta}{\sqrt{a^2\cos^2\theta+b^2\sin^2\theta}}\biggr)\times \notag\\ &\sqrt{a^2\cos^2\theta+b^2\sin^2\theta}~d\theta,\label{E_tension}
\end{align}
which is proportional to the area difference between the contact area (red in Fig.~\ref{uptakeprocess}) and the area of its projection.
The work done by the active particle for the adhered part is calculated as $E_f^{\rm ad}=-fb(1-{\rm cos}\alpha)$.
Given these and using $e=b/a$, the free energy of the wrapping part reads
\begin{align}
\frac{E_{\rm ad}^{\rm tot}}{\kappa}=&\int_0^{\alpha}\pi\sin\theta\biggl\{e^2\frac{[2+(e^2-1)\sin^2\theta]^2}{[1+(e^2-1)\sin^2\theta]^3} + \notag\\
& 2\frac{a^2}{\lambda^2}\biggl(1-\frac{\cos\theta}{\sqrt{\cos^2\theta+e^2\sin^2\theta}}\biggr) - 2\frac{\omega a^2}{\sigma\lambda^2}\biggr\}\times \notag\\
& \sqrt{\cos^2\theta+e^2\sin^2\theta}~d\theta - 2\pi\frac{aef}{\lambda f_0}(1-\cos\alpha),\label{E_Wrapping}
\end{align}
where $\lambda=\sqrt{\kappa/\sigma}$ and $f_{\rm 0}=2\pi\sqrt{\kappa\sigma}$ feature a typical length scale and a force scale, respectively.

For the free part of the membrane, its elastic energy comes from the axisymmetrically curved membrane shape described by $r(s)$, $z(s)$ and $\psi(s)$ [see Fig.~\ref{uptakeprocess}(b)], where $s$ is the arc length of the free membrane. The coordinates $r(s)$ and $z(s)$ depend on $\psi(s)$ through constraints $\dot{r}={\rm cos}\psi$ and $\dot{z}=-{\rm sin}\psi$, where the dots denote a derivative with respect to the arc length. The total energy of the free membrane, with the two principal curvatures given by $\dot{\psi}$ and $({\rm sin}\psi)/r$, can be written as~\cite{F.Julicher1994,M.Deserno2004,U.Seifert1990}
\begin{align}
\frac{E_{\rm free}^{\rm tot}}{\kappa}=\pi\int_{0}^{S}ds~\mathcal{L}(\psi,\dot{\psi},r,\dot{r},\dot{z},\eta,\xi,f),\label{E_tot_free}
\end{align}
where $\mathcal{L}$ is a Lagrangian defined by
\begin{align}
\mathcal{L}=~& r\biggl(\dot{\psi}+\frac{{\rm sin}\psi}{r}\biggr)^2+2\frac{\sigma}{\kappa}r(1-{\rm cos}\psi)-\frac{f}{\pi\kappa}{\rm sin}\psi \notag\\
& +\eta(\dot{r}-{\rm cos}\psi)+\xi(\dot{z}+{\rm sin}\psi).\label{Lagrangefunction}
\end{align}
Here $\eta(s)$ and $\xi(s)$ are Lagrangian multipliers used to impose the geometric constraints between $r$, $z$ and $\psi$. 
The term associated with the active force $f$ is proportional to the membrane height of the free part $Z_{\mathrm{free}}=\int_0^L \sin\psi ds $. 
A variation of the energy functional Eq.~(\ref{E_tot_free}) against the shape variables $r(s)$, $z(s)$ and $\psi(s)$ produces a set of shape equations, of which the details can be found in Appendix A. 
Here, we take the value of $\xi$ as a constant which equals to zero due to the fact that its first order derivative is zero, as well as the variation of the energy against $z(0)$ is zero (see Appendix A). 
With boundary conditions at the contact point between the particle and the membrane, $\psi(0)=\arctan(e\tan\alpha)$ for $\alpha\leq\pi/2$, $\psi(0)=\pi+\arctan(e\tan\alpha)$ for $\alpha>\pi/2$, and $r(0) = a\sin\alpha$, as well as $\psi(\infty) = 0$ and $z(\infty) = 0$ at the infinity, we numerically solve the shape equations for various $\alpha$ and obtain the total energy $E^{\mathrm{tot}}(\alpha)$ as a function of the wrapping angle $\alpha$. 
Based on the optimal wrapping angle $\alpha$ obtained via minimizing the total energy, we identify 4 types of wrapping states: nonwrapping (NW) when $\alpha=0$, small partial wrapping (SPW) when $0<\alpha\le \pi/2$, large partial wrapping (LPW) when $\pi/2\le\alpha<\pi$, and full wrapping (FW) when $\alpha = \pi$, as shown in Fig.~\ref{uptakeprocess}.
Here it should be noted that, in our theoretical model, we consider a special case when the pressure difference between the inside and outside of the plasma membrane, as well as the spontaneous curvature of the membrane, is neglected.

\section{RESULTS AND DISCUSSIONS}
\subsection{A. The effect of force, aspect ratio, and particle size on the wrapping states}
\begin{figure*}[htp]
  \includegraphics[width=\linewidth,keepaspectratio]{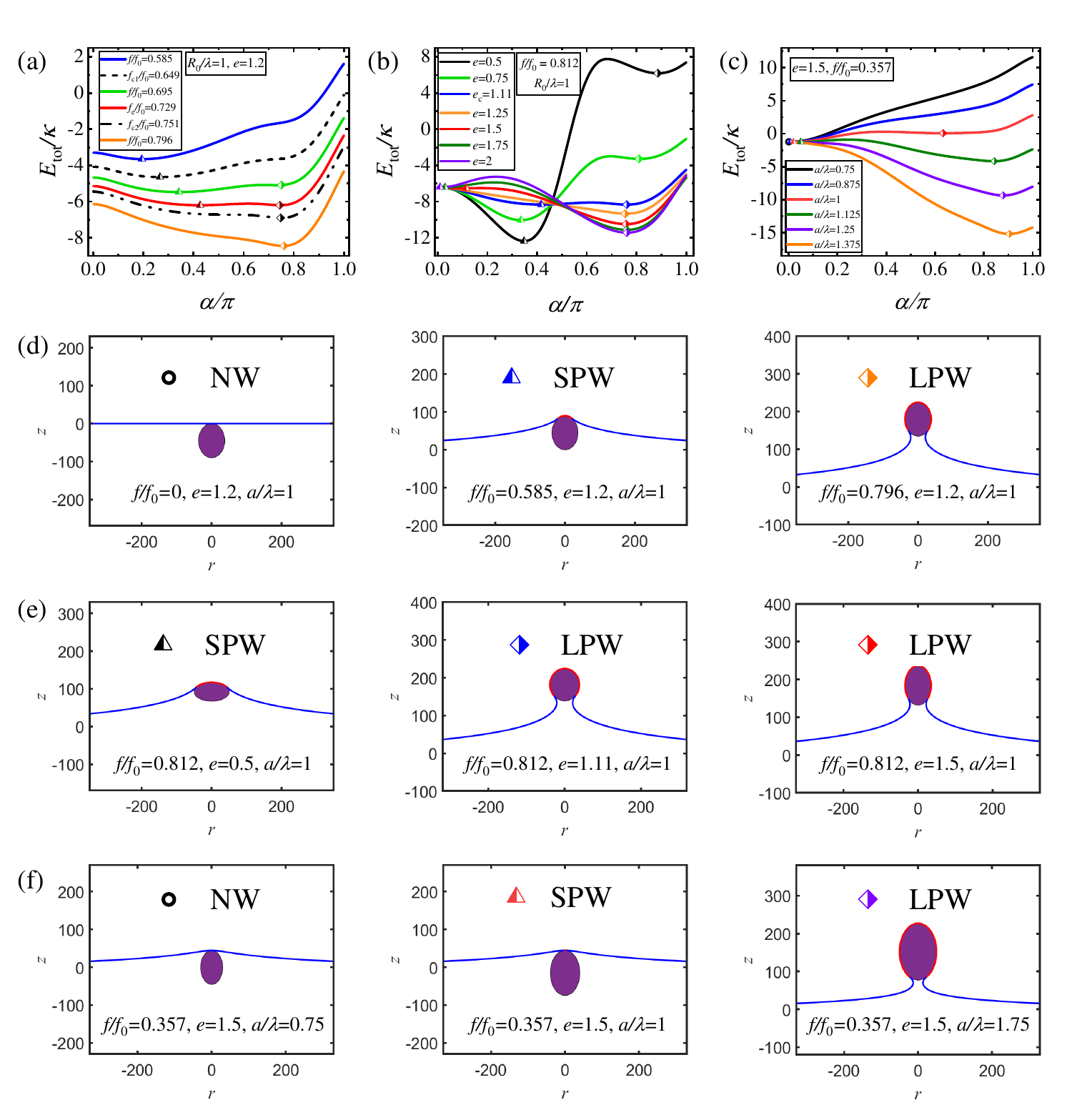}
  \caption{(Color online) Total free energy profile as a function of wrapping angle with $\omega=2\sigma$ for different (a) forces at fixed aspect ratio $e=1.2$ and particle size, (b) aspect ratios at fixed force $f/f_0=0.812$ and particle size, and (c) particle sizes at fixed aspect ratio $e=1.2$ and force $f/f_0=0.357$, respectively. The volume of the ellipsoidal particle is equal to that of a spherical particle with radius $R_0/\lambda=1$ if the particle size is not varied. Different wrapping states of the particle for various (d) forces, (e) aspect ratios, and (f) particle sizes corresponding to (a), (b), and (c) respectively.} \label{E Vs alpha}
\end{figure*}
As a first step of our model, in order to systematically evaluate the influences of the active force $f$ (the active protrusive force exerted by the self-propelled particle itself), the aspect ratio $e$ (ratio between the semimajor and semiminor axes), and the size of the particle on the wrapping degree at equilibrium, denoted by $\alpha/\pi$, we numerically solve the shape equations for different wrapping angles ranging from $0$ to $\pi$.
First, to examine the effect of the active force, we plot the total free energy profile $E_{\rm tot}(\alpha|f)$ as a function of the wrapping angle $\alpha$ for an ellipsoidal particle with aspect ratio $e=1.2$ at different forces $f$ under the condition of fixed particle volume ($V=4\pi ea^3/3=4\pi R_0^3/3$ with $R_0/\lambda=1.5$), as shown in Fig.~\ref{E Vs alpha}(a). Similar profiles can be found for $e< 1$ and $e=1$~\cite{Xiao2022}. The surface tension is set as $\sigma=0.012~k_BT/{\rm nm}^2$ unless specified.
In the presence of small active force [smaller than a threshold $f_{\rm c1}$], we can clearly see that there is only one stable wrapping state, i.e. the SPW state [see the blue and white triangle on the blue curve].
Further increase of $f$ leads to the occurrence of a metastable state, corresponding to the LPW state [see the green curve].
As the force increases to a critical value $f_{\rm e}$, the above two states becomes equal energetically with a barrier in between, indicating a first-order transition occurring anytime [see the red curve].
If the active force lies between $f_{\rm e}$ and another threshold $f_{\rm c2}$, the stable wrapping state will shift from SPW to LPW.
 If the force is larger than $f_{\rm c2}$, the metastable SPW state will disappear, leaving LPW the only stable wrapping state [see the orange curve]. Consequently, if $f_{\rm c1}<f<f_{\rm e}$, an SPW state is energetically more favorable, while if $f_{\rm e}<f<f_{\rm c2}$, a LPW state is more favorable.
If the active force $f\geq f_{\rm c2}$, the SPW-to-LPW energy barrier and the SPW state itself disappear. Such a double-well structure of free energy profile has been confirmed for spherical particles~\cite{Xiao2022}.

As a following step, in order to probe the effect of the aspect ratio $e$, we plot the total free energy against the wrapping angle for different aspect ratios $e$ at fixed active force $f$ and particle size ($V=4\pi ea^3/3=4\pi R_0^3/3$ with $R_0/\lambda=1$), as demonstrated in Fig.~\ref{E Vs alpha}(b). The dependence of wrapping state on aspect ratio reflects that there exist two local minima, with one corresponding to an SPW state and the other corresponding to a LPW state. The blue curve in Fig.~\ref{E Vs alpha}(b) demonstrates that there exists a critical aspect ratio $e_{\rm c}$ at which the SPW state and the LPW or the FW state have the same total free energy. If the aspect ratio decreases below the critical value $e_{\rm c}$, the stable state is an SPW one, as shown by the black and green curves.
Meanwhile the stable state will shift to a LPW one from an SPW one if the aspect ratio is larger than the critical value $e_{\rm c}$ [see the red, olive, and purple curves].

Furthermore, we elaborate upon the influence of particle size on the variation of the total free energy as a function of the wrapping angle $\alpha$ at fixed aspect ratio $e=1.5$ and active force, as demonstrated in Fig.~\ref{E Vs alpha}(c).
The curves in Fig.~\ref{E Vs alpha}(c) exhibit that only stable NW (SPW) state exists for small particle size $a/\lambda=0.75$ ($a/\lambda=0.875$) at $f/f_0=0.357$ [see black and blue curves].
A further increase of particle size gives rise to a metastable LPW state [see the red curve with particle size $a/\lambda=1$] besides the stable NW or SPW state. If the particle size goes beyond a threshold value, the metastable LPW state becomes a stable one [see olive curve].
If the particle size continues to increase, the metastable SPW state will finally vanish [see purple and orange curves].
Figures.~\ref{E Vs alpha}(d), (e) and (f) show 3 typical wrapping states at different combinations of active forces, aspect ratio, and particle size, respectively.

\subsection{B. Transition from SPW to LPW with hysteresis feature}
\begin{figure}[htp]
\centerline{\includegraphics[width=\linewidth,keepaspectratio]{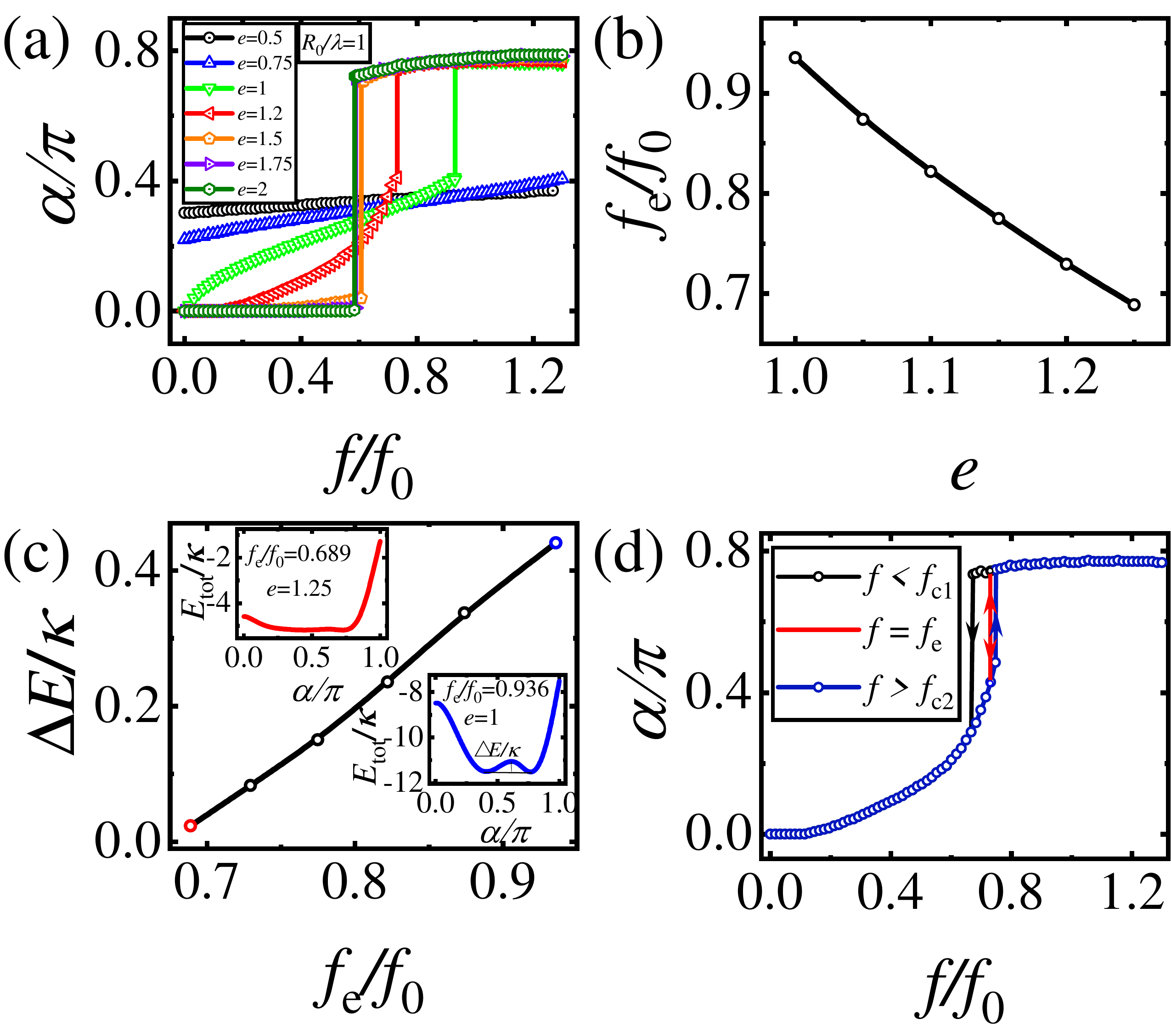}}
  \caption{(Color online) (a) The dependence of $\alpha$ on $f$ for different aspect ratios with $\omega=2\sigma$. (b) The dependence of critical force $f_{\rm e}$ on aspect ratio. (c) The energy barrier $\Delta E/\kappa$ separating the SPW and the LPW states against the critical force $f_{\rm e}$. (d) A typical hysteresis associated with $\alpha$ and $f$ triggering the transition when $e=1.2$, corresponding to Fig.~\ref{E Vs alpha}(a). \label{f Vs alpha_e}}
\end{figure}

In order to understand the wrapping behaviors in the regime $f_{\rm c1}\le f\le f_{\rm c2}$, we next plot the optimum wrapping angle $\alpha$ against active force $f$ for different aspect ratios at fixed particle volume ($V=4\pi ea^3/3=4\pi R_0^3/3$ with $R_0/\lambda=1$), as shown in Fig.~\ref{f Vs alpha_e}(a).
It is found that the optimum wrapping angle $\alpha$ shows a snapthrough at the critical value $f_{\rm e}$, indicating a first-order transition. Such a transition occurs only for a particle with an intermediate aspect ratio at a critical value $f_{\rm e}$ decreasing monotonically with aspect ratio, as shown in Fig.~\ref{f Vs alpha_e}(b).
Plotting the energy barrier $\Delta E/\kappa$ between SPW and LPW against the critical force $f_{\rm e}/f_0$ [Fig.~\ref{f Vs alpha_e}(c)] exhibits a nearly linear dependence with a positive slope.
 Hysteresis also features such a transition process, as show in Fig.~\ref{f Vs alpha_e}(d). If the active force $f$ is small, the system exhibits a stable SPW state, as shown by the blue curve in Fig.~\ref{E Vs alpha}(a). Further increase of $f$ results in the appearance of the metastable LPW state (green curve in Fig.~\ref{E Vs alpha}(a)), corresponding to $f_{\rm c1}$, followed by an equality of energy between SPW and LPW (red curve in Fig.~\ref{E Vs alpha}(a)), corresponding to the critical point $f_e$. If the active force $f$ is larger than the critical value $f_e$, the SPW state tends to remain as a metastable state until it disappears (dotted dash black curve in Fig.~\ref{E Vs alpha}(a)), corresponding to $f_{\rm c2}$, if the fluctuation is not large enough. A similar explanation for the hysteresis feature of wrapping transition for spherical particles can also be found in Ref.~\cite{Xiao2022}.

\begin{figure}[htp]
\centerline{\includegraphics[width=\linewidth,keepaspectratio]{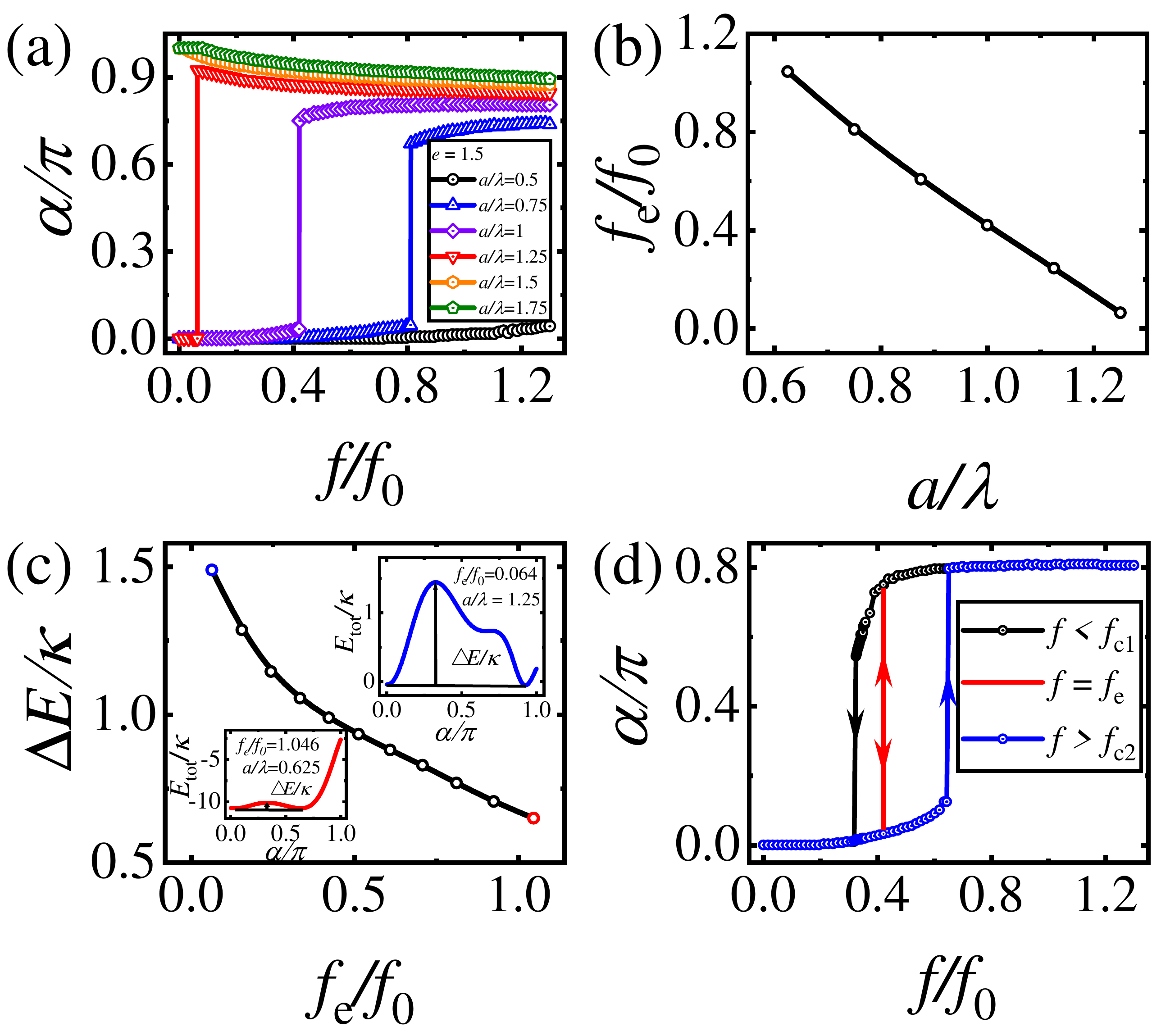}}
  \caption{(Color online) (a) The dependence of $\alpha$ on $f$ for different particle sizes with $\omega=2\sigma$. (b) The dependence of the critical force $f_{\rm e}$ on the particle size. (c) The energy barrier $\Delta E/\kappa$ between SPW and LPW against the critical force $f_{\rm e}$. (d) A typical hysteresis associated with the wrapping angle $\alpha$ and the active force $f$ triggering the transition. \label{f Vs alpha_a}}
\end{figure}

Similarly, the dependence of the optimum wrapping angle $\alpha$ on particle size is also investigated.
Figure~\ref{f Vs alpha_a}(a) depict the optimum wrapping angle $\alpha$ against active force $f$ for different particle sizes at fixed aspect ratio $e=1.5$, where the variation of the curves also indicate that the optimum wrapping angle $\alpha$ exhibits a sharp jump at the critical value $f_{\rm e}$, and such a first-order transition happens only for particles with intermediate sizes.
A further investigation reveals that such a critical value $f_{\rm e}$ decreases monotonically with the particle size, as shown in Fig.~\ref{f Vs alpha_a}(b).
On the other hand, the energy barrier $\Delta E/\kappa$ shows a remarkably decreasing behavior with the increase of critical force $f_{\rm e}$, a different behavior in comparison with Fig.~\ref{f Vs alpha_e}(c). In particular, when the force is small, for example $f_{\rm e}\approx 0.2f_0$ and given a typical value of $\kappa=20~k_BT$, the energy barrier for wrapping a particle can reach as high as $26~k_BT$, a value too large to be overcome by thermal fluctuations alone.
However, if $f_{\rm e}$ is large enough, the energy barrier is only about a few $k_BT$-s, a value close to the thermal fluctuation energy of membranes and just a tiny fraction of the membrane bending rigidity $\kappa$. Therefore the first-order transition is plausible, a conclusion consistent with the previous studies~\cite{S.Dasgupta2013,M.Deserno2004}.
In addition, a hysteresis that characterizes the transition process analogous to Fig.~\ref{f Vs alpha_e}(d) is founded as well [see Fig.~\ref{f Vs alpha_a}(d)].

\subsection{C. Phase diagram for force-induced wrapping behaviors}
\begin{figure*}[htp]
  \includegraphics[width=\linewidth,keepaspectratio]{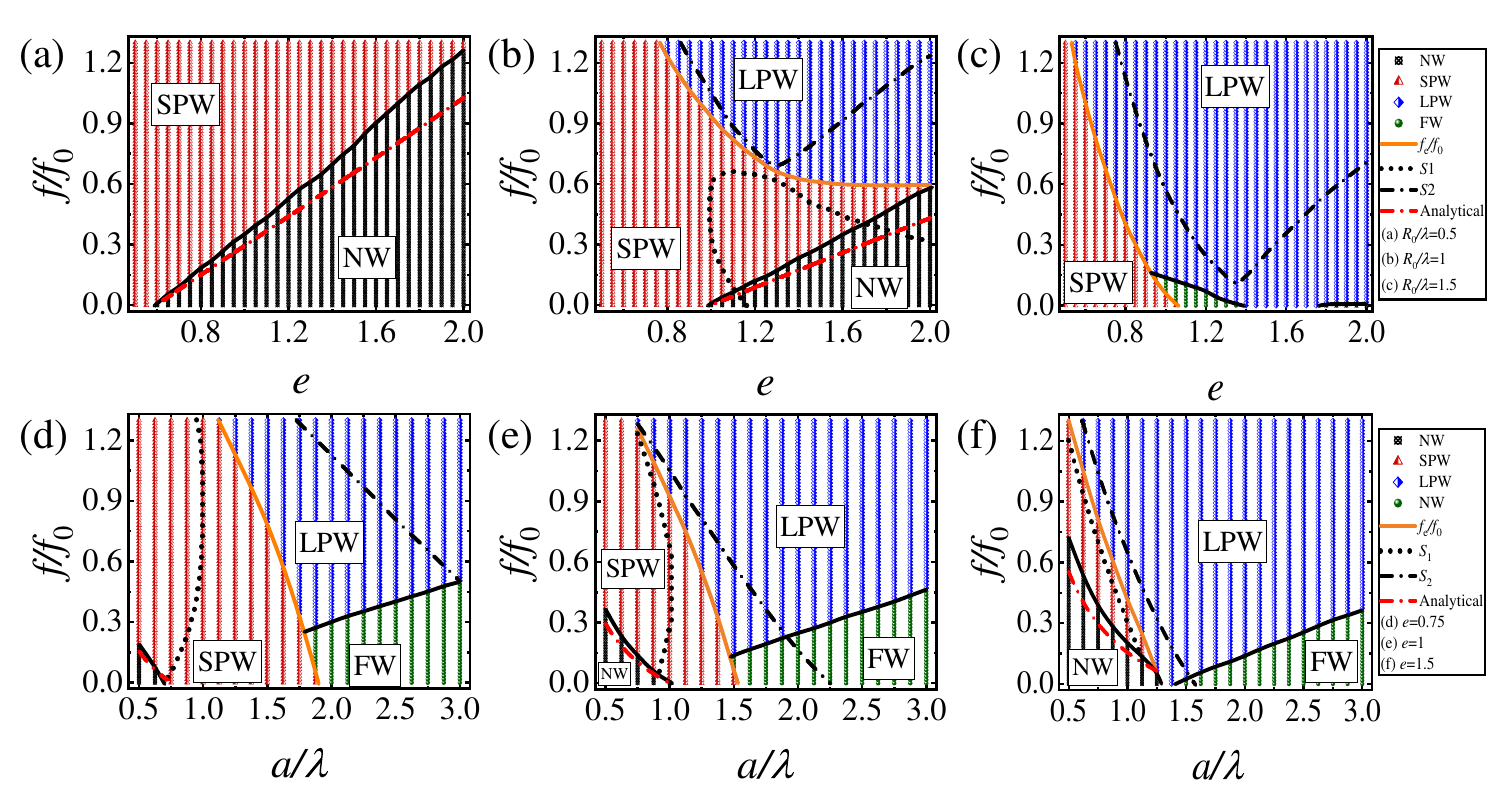}
  \caption{(Color online) Two-dimensional wrapping phase diagrams on the $f/f_0-e$ plane at fixed particle size (a) $R_0/\lambda=0.5$, (b) $R_0/\lambda=1$, and (c) $R_0/\lambda=1.5$ characterize the interrelated effects of active force and aspect ratio on the cellular uptake; Two-dimensional wrapping phase diagrams on the ($f/f_0-a/\lambda$) plane at fixed aspect ratios (d) $e=0.75$, (e) $e=1$, and (f) $e=1.5$ characterizing the interrelated effects of active force and aspect ratio on the cellular uptake. Where the ratio between the adhesion and tension strength is given by $\omega/\sigma=2$. The orange line indicates the discontinuous transition between SPW and LPW. The dotted line and the dash-dotted line indicate the spinodals accompanied with the transition. The black solid lines that separate NW and SPW, and LPW and FW indicate continuous second-order transitions. The red dashed line indicates the analytical solution to the boundary between NW and SPW. \label{ea Vs f PD}}
\end{figure*}

In order to systematically investigate how a wrapping state depends on active force $f$, aspect ratio $e$, particle size $a$ and membrane properties (i.e., adhesion strength $\omega$ and membrane tension $\sigma$), we first construct $f-e$ and $f-a/\lambda$ phase diagrams for different particle sizes and aspect ratios, respectively, as shown in Fig.~\ref{ea Vs f PD}, where four regions of different colors, corresponding to NW, SPW, LPW, and FW states respectively, can be identified.
A comparison among Figs.~\ref{ea Vs f PD}(a), (b), and (c) shows that the wrapping states enriches with the increase of particle size.
For instance, an LPW state emerges in Fig.~\ref{ea Vs f PD}(b) as compared with Fig.~\ref{ea Vs f PD}(a), and the FW state occurs in Fig.~\ref{ea Vs f PD}(c) as compared with Fig.~\ref{ea Vs f PD}(b).
Also, for small particle size, it is found that with the increase of active force, the wrapping degree is enhanced from NW to SPW (or from SPW to LPW) [see Figs.~\ref{ea Vs f PD}(a) and (b)].
Whereas the wrapping degree is reduced from FW to LPW for a large particle [see Fig.~\ref{ea Vs f PD}(c)].
Both of these transitions are continuous except for the first-order transition from SPW to LPW separated by the orange curves.
In addition, for a small aspect ratio, it is found that with the increase of active force, the wrapping degree is enhanced from NW to SPW, a transition that will be replaced by an NW-LPW one if the aspect ratio is large.
For Figs.~\ref{ea Vs f PD}(b) and (c), the dotted ($S_1$) and dash-dotted ($S_2$) curves represent the spinodals used to characterize the hysteresis features in Fig.~\ref{f Vs alpha_e}(d) and Fig.~\ref{f Vs alpha_a}(d).

To gain more insights into the effects of the particle size on the wrapping behaviors of the nonspherical active particle by a membrane, we construct the $f-a/\lambda$ phase diagrams for different aspect ratios, as shown in Figs.~\ref{ea Vs f PD}(d), (e), and (f).
With the increase of aspect ratio, both NW and LPW regimes are widened, and the SPW regime is contracted.
As an example, for the fixed aspect ratio $e=1.5$, it is found that with the increase of active force, the wrapping degree is enhanced from NW to SPW for small particles, but is reduced from FW to LPW for large particles. Both of these transitions are continuous.
 If the particle size falls into the intermediate range, increasing the active force leads to a discontinuous transition from SPW to LPW separated by the orange curves in Figs.~\ref{ea Vs f PD}(d), (e), and (f), with a sharp jump of the optimum wrapping angle across $\pi/2$. Hysteresis is found to feature the transition with its spinodals denoted as dotted ($S_1$) and dash-dotted ($S_2$) curves in Figs.~\ref{ea Vs f PD}(d), (e), and (f), respectively. Here it should be noted that LPW is a novel phase that does not exist in the absence of force.
In particular, from NW to SPW, as the membrane is just slightly deformed and remains almost flat ($\psi\ll 1$), it is reasonable to linearize the shape equations. The obtained analytical expression for the boundary curves between NW and SPW, is supported by the numerical results, as shown by the red dashed line in Fig.~\ref{ea Vs f PD}. The detailed derivation of such an expression can be found in Appendix B.

\begin{figure}[htp]
  \includegraphics[width=\linewidth,keepaspectratio]{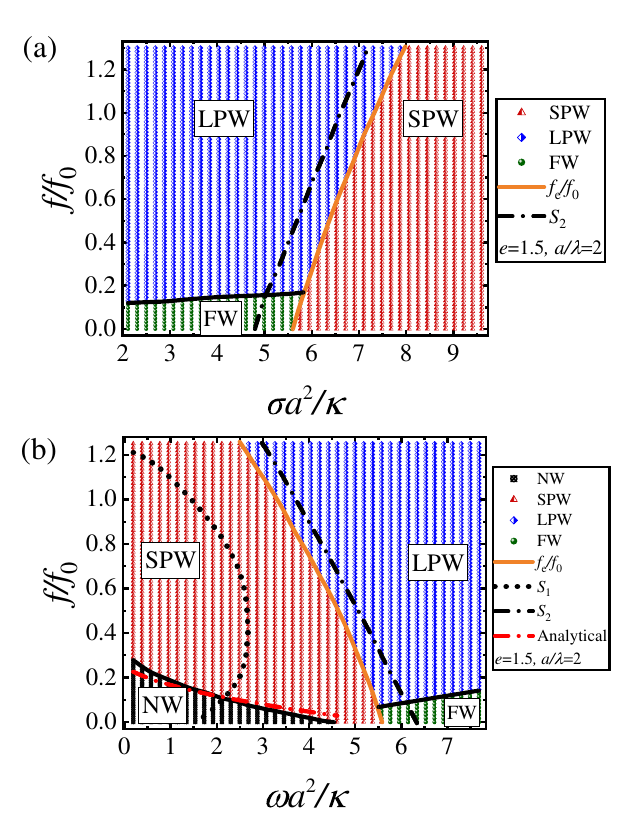}
  \caption{(Color online) Two-dimensional wrapping phase diagrams in the projection of (a) $f/f_0-\omega a^2/\kappa$ and (b) $f/f_0-\sigma a^2/\kappa$ planes characterizing the interrelated effects of active force and adhesion energy density, and active force and membrane tension on the cellular uptake, where the aspect ratio and particle size are fixed as 1.5 and 2, respectively. The orange line indicates the discontinuous transition between SPW and LPW. The dotted line and the dash-dotted line indicate the spinodals accompanied with the transition. The black solid lines that separate NW and SPW, and LPW and FW indicate continuous transitions. The red dashed line demonstrates the analytical solution to the boundary between NW and SPW. \label{wsigma Vs f PD}}
\end{figure}

Finally, to further reveal the effect of membrane properties (including the adhesion energy density and the membrane tension) on the wrapping behaviors of the nonspherical active particle by a membrane, we also construct phase diagrams in the $f-\omega$ and the $f-\sigma$ planes, respectively, as shown in Fig.~\ref{wsigma Vs f PD}.
It is found that weak adhesion force of the membrane leads to an enhancing wrapping degree (from NW to SPW) with the increase of active force (Fig.~\ref{wsigma Vs f PD}(b)). In low active force, and strong adhesion or loose membrane regime, it is possible that strong membrane adhesion dominates the wrapping over surface tension and elasticity of the membrane, resulting in an FW state. The increase of active force tends to lift up the membrane, pushing the part of the membrane around the particle into a tube-like shape. This in turn reduces the wrapping degree and consequently leads to a transition from an FW state to an LPW one.
In addition, if the membrane adhesion energy density and the membrane tension are located in an intermediate range, a discontinuous first-order transition from SPW to LPW separated by the orange curves in Fig.~\ref{wsigma Vs f PD} can be triggered by increasing the active force.
Moreover, the dotted ($S_1$) and dash-dotted ($S_2$) curves in the dotted ($S_1$) and dash-dotted ($S_2$) curves denote the spinodals used to characterize the hysteresis features, which are also plotted in Fig.~\ref{wsigma Vs f PD}.
Based on the detailed derivation in Appendix B, we also plot the analytical boundary curves separating NW and SPW [see red dashed lines in Fig.~\ref{wsigma Vs f PD}], demonstrating a good agreement with the numerical results.

\section{Discussion}
\subsection{Competition among bending energy, adhesive energy and the work done by the force}
\begin{figure*}[htp]
  \includegraphics[width=\linewidth,keepaspectratio]{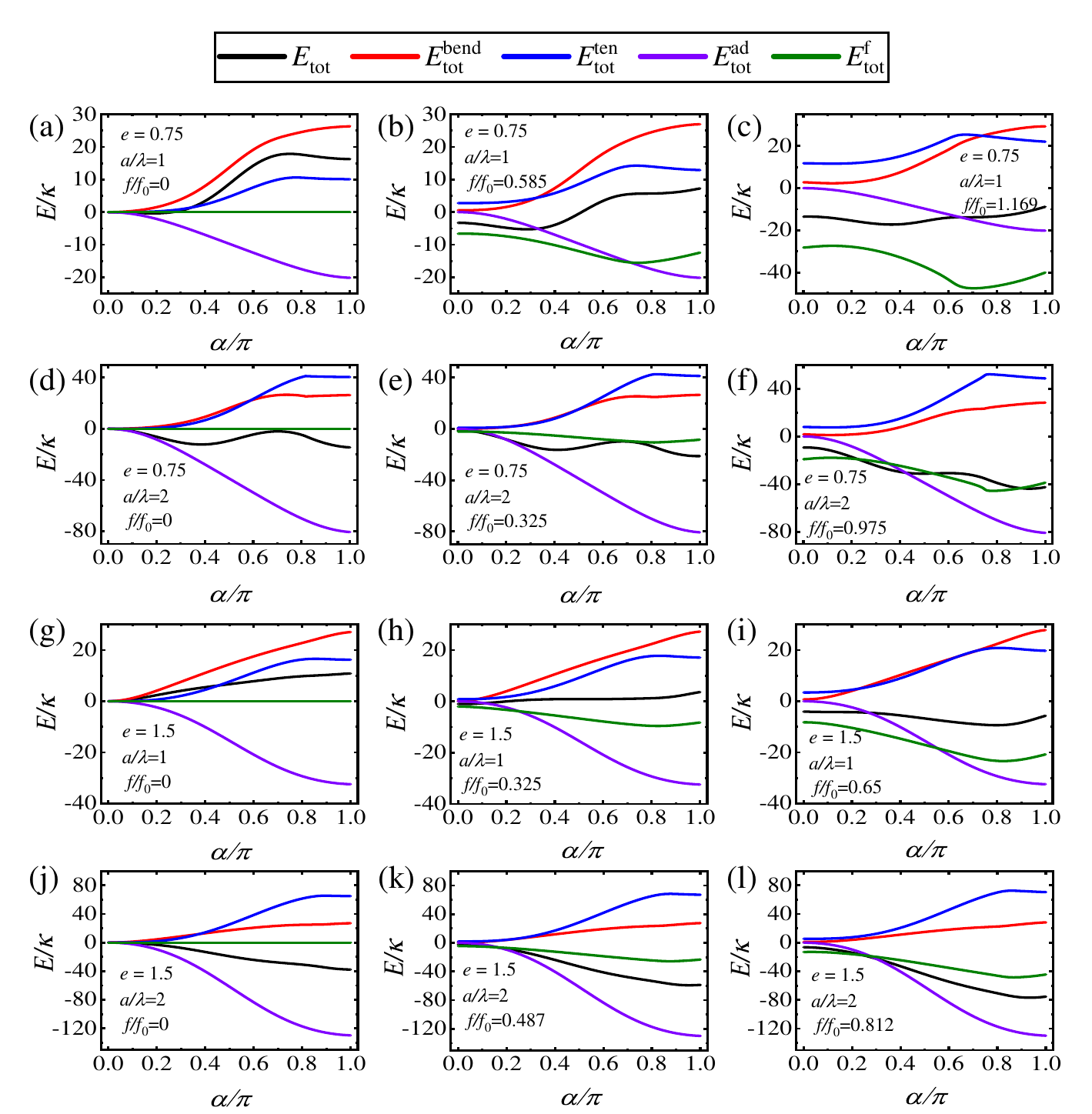}
  \caption{(Color online) Energy profile as a function of the wrapping angle for different active forces with $\omega=2\sigma$. The aspect ratio and particle size are set as (a)-(c) $e=0.75$ and $a/\lambda=1$, (d)-(f) $e=0.75$ and $a/\lambda=2$, (g)-(i) $e=1.5$ and $R/\lambda=1$, and (j)-(l) $e=1.5$ and $R/\lambda=2$. \label{E_alpha}}
\end{figure*}
We have studied the shape transformations of a flat membrane generated by a self-propelled nonspherical NP, and shown that a novel phase of LPW emerges as a result of the force. The physics behind the transitions between different wrapping states comes from the the competition among the elastic energy (consisting of bending energy and tension energy), adhesive energy, and the work done by the active force. The calculated total energy, elastics energy (including bending energy and tension energy), adhesion energy and the work done by active force as a function of wrapping angle shown in Fig.~\ref{E_alpha} demonstrate that the wrapping effect is governed by a balance among these energy players.
In the absence of force, for small aspect ratio and small particle, adhesion-induced wrapping cannot compensate the high energy cost of bending, and therefore a NW state is the most stable one. However, the introduction of the work done by the active force reduces the total energy, which enables the membrane to deform and wrap around the particle, even though the wrapping is partial and small due to the little contribution made by adhesion. In contrast, for large aspect ratio and large particle ($R\gg \lambda$) at small active force, the FW state is the most stable one because the penalty of elastic energy is sufficiently balanced by adhesive energy. Increasing the force tends to lift up the membrane, which in turn reduces the wrapping degree and consequently leads to an LPW state.
According to the total free energy given by Eq.~(\ref{HelfrichFE}), apart form the bending energy, membrane deformations are mainly opposed by tension.
Therefore, if the particle without activity (which means it does not have active force), for weak adhesion strength and high membrane tension, the positive adhesion energy is unable to drive the wrapping of particle by the membrane. In this case, some external forces are required for the activation of wrapping.
While for strong adhesion force and low membrane tension, the adhesion energy is sufficient to drive the wrapping of particle by membrane.
In this paper, we choose the aspect ratio and particle size such that in the absence of force, increasing (reducing) the adhesion strength $\omega$ (membrane tension $\sigma$) would lead to a transition from SPW (FW) state to FW (SPW) state.
For a cell to engulf a self-propelled nonspherical particle, an FW wrapping state is necessary to enclose the particle inside a vesicle. The phase diagrams shown in Figs.~\ref{ea Vs f PD} and ~\ref{wsigma Vs f PD} suggest that if the particle activity is very strong, it should be difficult for the cell to engulf a very large particle.

\section{Conclusion}
In summary, based on the total energy functional, we study the wrapping states of a self-propelled nonspherical particle when it is pushed against a membrane. It is found that the active force generated by the particle is able to trigger a first-order wrapping transition, accompanied with a hysteresis behavior. Such a transition provides a deeper insight into the wrapping behaviors induced by a self-propelled nonspherical particle. The wrapping states of the active particle are tunable by active force, aspect ratio, particle size, and membrane properties (including the adhesion energy density and the membrane tension), as demonstrated by the phase diagrams in the two-parameter space. It is also identified that the wrapping degree can be enhanced (for small particle, weak adhesion strength, and high membrane tension) or decreased (for large particle, strong adhesion strength, and low membrane tension) upon increasing the active force of the particle. Our results provide a useful guidance for engineering active particle-based therapeutics to promote biomedical applications.

\section{ACKNOWLEDGMENTS}
We acknowledge financial support from National Natural Science Foundation of China under Grant Nos.12147142, 11974292, 12174323, and 1200040838, and 111 project B16029.

\section{Appendix A: Derivation of the membrane shape equations}
For axisymmetric surfaces, from the energy functional Eq.~(\ref{E_tot_free}) in the main text and by variational methods one can derive the Euler-Lagrange equations
\begin{align}
\ddot{\psi}=~&-\frac{{\rm cos}\psi}{r}\dot{\psi}+\frac{{\rm sin}\psi{\rm cos}\psi}{r^2}+\frac{\sigma}{\kappa}{\rm sin}\psi-\frac{f}{2\pi\kappa r}{\rm cos}\psi \notag\\
&+\frac{\eta}{2r}{\rm sin}\psi+\frac{\xi}{2r}{\rm cos}\psi,\label{SI:ddpsi}
\end{align}
\begin{align}
\dot{\eta}=\dot{\psi}^2-\frac{{\rm sin}^2\psi}{r^2}+2\frac{\sigma}{\kappa}(1-{\rm cos}\psi),\label{SI:deta}
\end{align}
\begin{align}
\dot{\xi}=0,\label{SI:dxi}
\end{align}
\begin{align}
\dot{r}={\rm cos}\psi,\label{SI:dr}
\end{align}
\begin{align}
\dot{z}=-{\rm sin}\psi.\label{SI:dZ}
\end{align}
In order to numerically solve the above equations, we first map the region $s\in[0,\infty]$ to a finite region $s\in[0,S_{tot}]$ and introduce a parameter $u=s/S_{tot}$ which is defined on a fixed interval $[0, 1]$. All the functions of $s$ are therefore transformed into functions of $u$. The derivative $\frac{d}{ds}$ are replaced with $\frac{1}{S_{tot}}\frac{d}{ds}$ and the five equations~(\ref{SI:ddpsi})-(\ref{SI:dZ}) are transformed into ordinary differential equations with respect to the parameter $u$. The equations are all first order except Eq.~(\ref{SI:ddpsi}), which is second order of $\psi$. They are equivalent to 6 first order ordinary differential equations. In addition, with the unknown parameter $S_{tot}$, we need 7 boundary conditions to complete the problem. These boundary conditions include: $\psi(u=0)=\arctan(e\tan\alpha)$ for $\alpha\leq\pi/2$, $\psi(u=0)=\pi+\arctan(e\tan\alpha)$ for $\alpha>\pi/2$, $\psi(u=1)=0$, $r(u=0)=a\sin\alpha$, $r(u=1)=R_b$, $\xi(0)=0$, $z(u=1)=0$. In practice, we let $R_b$ to be a large enough number such that the results are not changed for values greater than $R_b$. Here we only have 6 boundary conditions and one more boundary condition still needed.
To complete the boundary conditions, we consider a homogeneous membrane, so that the Lagrangian $\mathcal{L}$ is explicitly independent of the arc length $s$. As a result, the Hamiltonian $\mathcal{H}\equiv -\mathcal{L}+\dot{\psi}\partial\mathcal{L}/\partial\dot{\psi}+\dot{r}\partial\mathcal{L}/\partial\dot{r}+\dot{z}\partial\mathcal{L}/\partial\dot{z}$ is a conserved quantity~\cite{F.Julicher1994} given by
\begin{align}
\mathcal{H}=~& r\biggl(\dot{\psi}^2-\frac{{\rm sin}^2\psi}{r^2}\biggr)-2\frac{\sigma}{\kappa}r(1-{\rm cos}\psi)+\frac{f}{\pi\kappa}{\rm sin}\psi \notag\\
& +\eta{\rm cos}\psi-\xi{\rm sin}\psi.\label{Hamitonian}
\end{align}
Due to that $\mathcal{H}$ is conserved along the arc length, i.e., $\mathcal{H}(s)=0$. We therefore impose the seventh boundary condition which is $\mathcal{H}(u=0)=0$. The 6 first order equations with an unknown parameter $S_{tot}$ plus 7 boundary conditions constitute a well-defined boundary value problem (BVP) that can be numerically solved by the Matlab solver 'bvp4c'.

By combining Eqs.~(\ref{SI:ddpsi}) and~(\ref{Hamitonian}), and letting $\xi=0$, one can derive the general shape equation
\begin{align}
\ddot{\psi}r^2\cos\psi + \dot{\psi}r\cos^2\psi &+ \frac{1}{2}\dot{\psi}^2r^2\sin\psi-\frac{1}{2}(\cos^2\psi+1)\sin\psi \notag\\
&-\frac{\sigma}{\kappa}r^2\sin\psi + \frac{fr}{2\pi\kappa}=0.\label{SecondShapeEquation}
\end{align}

\section{Appendix B: Analytical expression for the critical curve that separates NW and SPW}
As mentioned in the main text, the total free energy of the system can be divided into two main parts: the wrapping part of the membrane in contact with the particle, and the free part of the membrane.
At the NW-SPW transition, the wrapping angle is zero, $\alpha=0$. According to the local mean curvature on the contact point Eq.~(\ref{Meancurvature}), we assume that the particle's local effective radius at this point is
\begin{align}
R_{\rm eff} = \frac{1}{H\mid_{\theta=0}}=\frac{a^2}{b}=\frac{a}{e}.\label{Reff}
\end{align}
As a result, the contribution of the wrapping part to the total free energy can be rewritten as
\begin{align}
\frac{E_{\rm ad}^{\rm tot}}{\kappa}=\pi(1-{\rm cos}\alpha)\biggl[\frac{R_{\rm eff}^2}{\lambda^2}(1-{\rm cos}\alpha)-2\frac{\omega R_{\rm eff}^2}{\sigma \lambda^2}-\frac{2R_{\rm eff}f}{\lambda f_{\rm 0}}+4\biggr].\label{E_Wrappingeff}
\end{align}
Here, it should be noted that we have assumed the wrapping part as part of a sphere of an effective radius $R_{\rm eff}$.

For weakly deformed membrane ($\psi\ll 1$) and small value of $\alpha$, Eq.~(\ref{SecondShapeEquation}) can be linearized as
\begin{align}
\ddot{\psi}r^2 + \dot{\psi}r - (1+\lambda^{-2}r^2)\psi = -\frac{fr}{2\pi\kappa}.\label{LinearizedShapeEquation}
\end{align}
 The small value of function $\psi$ leads to an proper approximation that the radial coordinate $r$ equals to the arc length $s$ to the first order due to $dr = ds \cos\psi \approx ds + O(\psi^2)$. Given this, the general solution to Eq.~(\ref{LinearizedShapeEquation}) reads
\begin{align}
\psi=\frac{f\lambda^2}{2\pi\kappa r} + A\textit{I}_1(r/\lambda) + B\textit{K}_1(r/\lambda),\label{Generalsolution}
\end{align}
where $\textit{I}_1(x)$ and $\textit{K}_1(x)$ are first-order modified Bessel functions, and $A$ and $B$ are integration constants. According to the boundary conditions $\psi(r = R_{\rm eff}\sin\alpha) = \alpha$ and $\psi(r = +\infty) = 0$, one can determine that $A=0$ and $B=[\alpha-f\lambda^2/(2\pi\kappa R_{\rm eff}\sin\alpha)]/K_1(R_{\rm eff}\sin\alpha/\lambda)$. Therefore, in the limit of $\alpha\ll 1$ and $\psi\ll 1$, we can calculate the work done by the active particle for the free part as
\begin{align}
E_{\rm f}^{\rm free}&= -f\int_{R_{\rm eff}\alpha}^{R_{\rm b}} \psi {\rm d}r \notag\\
&=-\frac{f\lambda(f\lambda^2-2\pi\kappa R_{\rm eff}\alpha^2)}{2\pi\kappa R_{\rm eff}\alpha K_1(R_{\rm eff}\alpha/\lambda)}[K_0(R_{\rm b}/\lambda)-K_0(R_{\rm eff}\alpha/\lambda)] \notag\\
&+\frac{f^2\lambda^2}{2\pi\kappa}{\rm ln}(\frac{R_{\rm eff}\alpha}{R_{\rm b}}).\label{EFfree}
\end{align}
\begin{figure}[htp]
  \includegraphics[width=\linewidth,keepaspectratio]{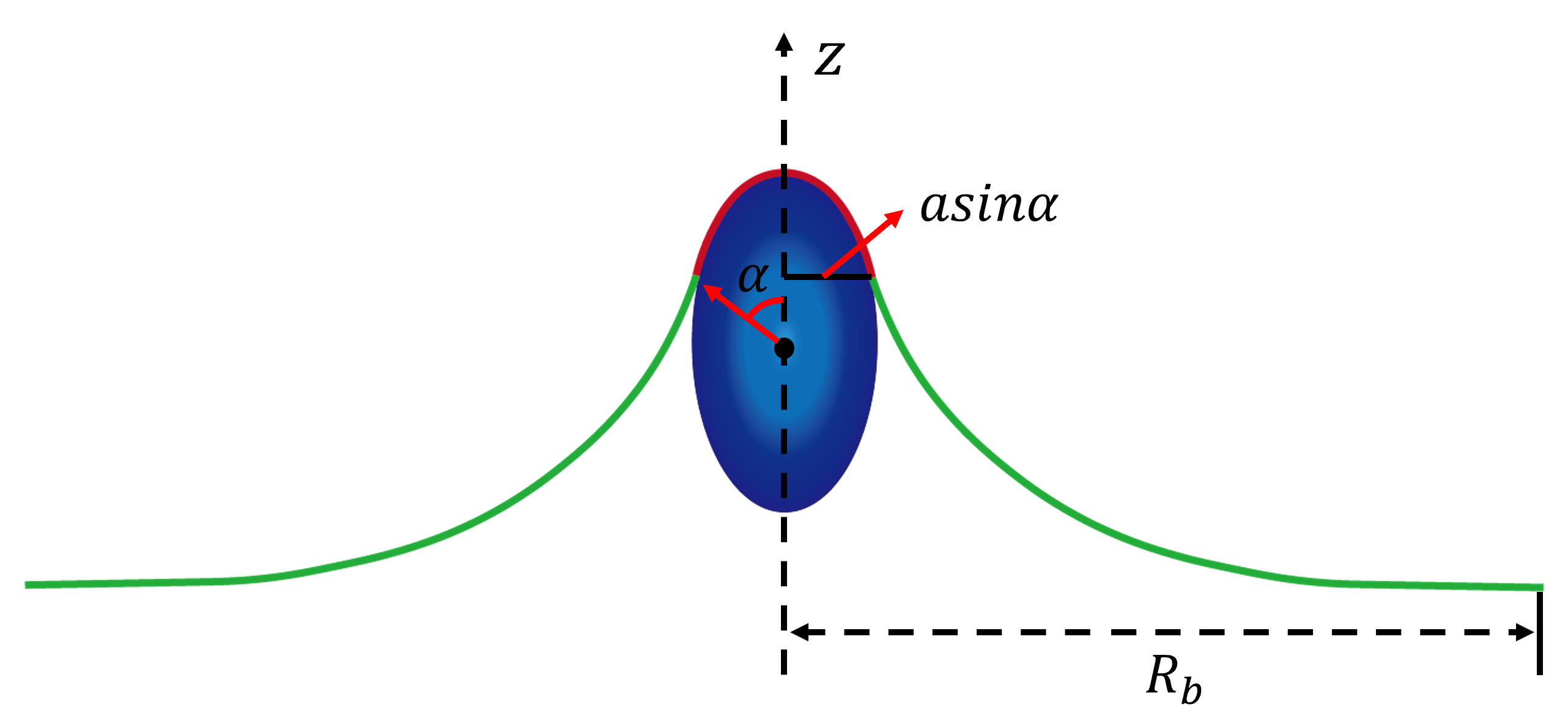}
  \caption{(Color online) Definition of some geometric parameters. \label{Fig8}}
\end{figure}
Here the lower limit for the integration variable $r$ is $R_{\rm eff}\sin\alpha \approx R_{\rm eff}\alpha$, as shown in the schematic Fig.~\ref{Fig8}. The upper limit $R_{\mathrm{b}}$ in practice is chosen to be a finite value so as to avoid the divergence of the integral Eq.~(\ref{EFfree}) when $R_{\mathrm{b}} \rightarrow \infty$, but large enough (e.g. $R_{\mathrm{b}}/\lambda=1000$) so that further increasing $R_{\mathrm{b}}$ brings little change to the result when calculating the second derivative of the total energy, as discussed later in this subsection. The bending energy of the free part of the membrane reads
\begin{align}
E_{\rm bend}^{\rm free}&= \pi\kappa\int_{R_{\rm eff}\alpha}^{R_{\rm b}}\biggl(\dot{\psi}+\frac{\psi}{r}\biggr)^2 r {\rm d}r \notag\\
&=\biggl(\frac{f\lambda^2 - 2\pi\kappa R_{\rm eff}\alpha^2}{2\sqrt{2\pi\kappa}R_{\rm eff}\alpha\lambda K_1(R_{\rm eff}\alpha/\lambda)}\biggr)^2 \biggl[R_{\rm b}^2(K_0^2(R_{\rm b}/\lambda) \notag\\
&-K_1^2(R_{\rm b}/\lambda))+R_{\rm eff}^2\alpha^2(K_1^2(R_{\rm eff}\alpha/\lambda)-K_0^2(R_{\rm eff}\alpha/\lambda))\biggr],\label{Ebfree}
\end{align}
and the tension energy is given by
\begin{align}
E_{\rm tens}^{\rm free}&= \pi\sigma\int_{R_{\rm eff}\alpha}^{R_{\rm b}} r\psi^2 {\rm d}r \notag\\
&=\frac{\sigma}{8\pi\kappa^2\alpha^2 R_{\rm eff}^2K_1^2(R_{\rm eff}\alpha/\lambda)} \biggl\{(f\lambda^2-2\pi\kappa R_{\rm eff}\alpha^2)^2  \notag\\
&\biggl[R_{\rm b}^2\bigl(K_1^2(R_{\rm b}/\lambda)-K_0^2(R_{\rm b}/\lambda)\bigr) +R_{\rm eff}^2\alpha^2\bigl(K_0^2(R_{\rm eff}\alpha/\lambda) \notag\\
&-K_1^2(R_{\rm eff}\alpha/\lambda)\bigr) \biggr] - 2\lambda(2\pi\kappa R_{\rm eff}\alpha^2-f\lambda^2)K_0(R_{\rm b}/\lambda)\cdot \notag\\
&\biggl[R_{\rm b}(2\pi\kappa R_{\rm eff}\alpha^2-f\lambda^2)K_1(R_{\rm b}/\lambda)+2fR_{\rm eff}\alpha\lambda^2K_1(R_{\rm eff}\alpha/\lambda) \biggr] \notag\\
&+2R_{\rm eff}\alpha(4\pi^2\kappa^2R_{\rm eff}^2\alpha^4\lambda-f^2\lambda^5)K_0(R_{\rm eff}\alpha/\lambda)K_1(R_{\rm eff}\alpha/\lambda) \notag\\
&+2f^2R_{\rm eff}^2\alpha^2\lambda^4K_1^2(R_{\rm eff}\alpha/\lambda){\rm ln}\frac{R_{\rm b}}{R_{\rm eff}\alpha} \biggr\}.\label{Etfree}
\end{align}
Summing these three terms and the total energy for the adhesion part, and doing a Taylor expansion with respect to $\alpha$ to the second order of $\alpha$, leads to
\begin{widetext}
\begin{align}
E_{\rm tot}/\kappa=&-\pi\left(\frac{f}{f_0}\right)^2\biggl[\frac{ R_{\rm b}}{\lambda}K_0(R_{\rm b}/\lambda)K_1(R_{\rm b}/\lambda)+{\rm ln}(R_{\rm b}/\lambda)+\gamma-{\rm ln2}\biggr] \notag\\
&+ \frac{\pi}{2}\left(\frac{f}{f_0}\right)^2\left(\frac{R_{\rm eff}}{\lambda}\right)^2\alpha^2\biggl[\left({\rm ln\alpha}+\frac{R_{\rm b}}{\lambda}K_0(R_{\rm b}/\lambda)K_1(R_{\rm b}/\lambda)+\frac{2}{(f/f_0)(R/\lambda)}+{\rm ln}(R_{\rm eff}/\lambda)+\gamma-{\rm ln2}-\frac{1}{2}\right)^2 \notag\\
&-\biggl(\frac{R_{\rm b}}{\lambda}K_0(R_{\rm b}/\lambda)K_1(R_{\rm b}/\lambda)\biggr)^2-2\frac{\omega/\sigma }{(f/f_0)^2}+\frac{1}{4}\biggr], \label{Etotfree+Etotad}
\end{align}
\end{widetext}
where $\gamma$ is the Euler Gamma function.
\begin{widetext}
\begin{align}
\frac{{\rm d}(E_{\rm tot}/\kappa)}{{\rm d}\alpha}=&\pi\left(\frac{f}{f_0}\right)^2\left(\frac{R_{\rm eff}}{\lambda}\right)^2\alpha\biggl[2\frac{R_{\rm b}}{\lambda}K_0(R_{\rm b}/\lambda)K_1(R_{\rm b}/\lambda)\left({\rm ln}\alpha + \frac{2}{(f/f_0)(R_{\rm eff}/\lambda)}+{\rm ln}(R_{\rm eff}/\lambda)+\gamma-{\rm ln}2\right) - 2\frac{\omega/\sigma}{(f/f_0)^2} \notag\\
&+\left(\frac{2}{(f/f_0)(R_{\rm eff}/\lambda)}+\gamma\right)^2 + \left({\rm ln}\alpha + \frac{4}{(f/f_0)(R_{\rm eff}/\lambda)}+{\rm ln}(R_{\rm eff}/\lambda)+2\gamma-{\rm ln}2\right)\biggl({\rm ln}\alpha+{\rm ln}(R_{\rm eff}/\lambda)-{\rm ln}2\biggr)
\biggr], \label{dEtot/dalpha}
\end{align}
\end{widetext}
The second order derivative of the total energy with respect to $\alpha$ is obtained as
\begin{widetext}
\begin{align}
\frac{{\rm d^2}(E_{\rm tot}/\kappa)}{{\rm d}\alpha^2}=&\pi\left(\frac{f}{f_0}\right)^2\left(\frac{R_{\rm eff}}{\lambda}\right)^2\biggl[\left(\frac{2}{(f/f_0)(R_{\rm eff}/\lambda)}+\gamma\right)
\left(\frac{2}{(f/f_0)(R_{\rm eff}/\lambda)}+\gamma+2\right)-2\frac{\omega/\sigma}{(f/f_0)^2} \notag\\
&+2\frac{R_{\rm b}}{\lambda}K_0(R_{\rm b}/\lambda)K_1(R_{\rm b}/\lambda)\left({\rm ln}\alpha + \frac{2}{(f/f_0)(R_{\rm eff}/\lambda)}+{\rm ln}(R_{\rm eff}/\lambda)+\gamma+1-{\rm ln}2\right) \notag\\
&+\biggl({\rm ln}\alpha+{\rm ln}(R_{\rm eff}/\lambda)-{\rm ln}2\biggr)^2+2\biggl({\rm ln}\alpha+{\rm ln}(R_{\rm eff}/\lambda)-{\rm ln}2\biggr)\biggl(\frac{2}{(f/f_0)(R_{\rm eff}/\lambda)}+\gamma+1\biggr)
\biggr]. \label{ddEtot/dalpha}
\end{align}
\end{widetext}
By setting ${\rm d^2}(E_{\rm tot}/\kappa)/{\rm d}\alpha^2=0$, we can get the analytical solution corresponding to the critical transition line between NW and SPW, which is shown by the red dash line in Figs.~\ref{ea Vs f PD} and ~\ref{wsigma Vs f PD} in the main text. A comparison between the analytical results and the exact numerical results indicates that the approximate expression is remarkably accurate.

\end{document}